\documentclass[aip,apl,reprint]{revtex4-1}
\usepackage{epsfig}
\begin{document}

%\preprint{}

\title{High Saturation Power Josephson Parametric Amplifier with GHz Bandwidth} %Title of paper



\author{O. Naaman}
\email[]{ofer.naaman@ngc.com}
\author{D. G. Ferguson}
\author{R. J. Epstein}

%\thanks{}
\affiliation{Northrop Grumman Systems Corp., Baltimore, Maryland 21240, USA }


\date{\today}

\begin{abstract}
We present design and simulation of a Josephson parametric amplifier with bandwidth exceeding 1.6 GHz, and with high saturation power approaching -90 dBm at a gain of 22.8 dB. An improvement by a factor of roughly 50 in bandwidth over the state of the art is achieved by using well-established impedance matching techniques. An improvement by a factor of roughly 100 in saturation power over the state of the art is achieved by implementing the Josephson nonlinear element as an array of rf-SQUIDs with a total of 40 junctions. WRSpice simulations of the circuit are in excellent agreement with the calculated gain and saturation characteristics.
\end{abstract}


\maketitle %\maketitle must follow title, authors, abstract and \pacs


Josephson parametric amplifiers have been in extensive use over the past few years, providing quantum limited noise performance at gains exceeding 20 dB, and enabling high fidelity qubit readout, \cite{Vijay11,Jeffrey14,Abdo14} squeezed microwave field generation,\cite{Murch13} weak measurement, \cite{Murch13b} and feedback control.\cite{Vijay12} However, state-of-the-art devices of the eponymous JPA\cite{Mutus13} and Josephson Parametric Converter (JPC) types\cite{Abdo17} suffer from either narrow bandwidth $\sim10$~MHz, or low saturation power $\sim-110$~dBm, or in many cases from both. Traveling-wave parametric amplifier \cite{Macklin15,White15} architectures can achieve large bandwidths of several GHz at the cost of high junction counts, typically exceeding 2000, with only modest improvement in saturation power. Here, we describe a flux-pumped JPA-type three-wave mixing amplifier with over 20 dB gain, in which we implement well-established impedance matching techniques to achieve over 1.6 GHz bandwidth, and a recently developed junction array design \cite{Naaman17} to achieve high saturation power approaching -90 dBm. Overall, this work represents 50-fold improvement over the state of the art in bandwidth and 100-fold improvement saturation power, in a circuit with less than 100 junctions. We compare the calculated amplifier response to Spice simulations of the full nonlinear circuit.

Both JPA and JPC, and their variants, are built with resonant structures embedding Josephson junctions or SQUIDs, which serve as the nonlinear active elements in the amplifier. Traditionally, their design has been driven by the principle that the loaded quality factor of the resonated nonlinearity must be relatively high in order to achieve high power gains. As a result, most Josephson parametric amplifiers are extremely narrow-band, and a considerable effort has been directed into making their center frequency tunable \cite{Roch12} to enable their practical use in the lab. This concept has been challenged recently by the work of Mutus \textit{et al.}\cite{Mutus14} and Roy \textit{et al.},\cite{Roy15} who have shown that JPAs have been traditionally operating far from their maximum possible gain-bandwidth product, and that high gains and wide-band operation can be achieved simultaneously by improving the amplifier impedance match to the 50 $\Omega$ environment.

The pumped nonlinearity in a parametric amplifier presents the signal port with an effective negative resistance, giving rise to reflection gain. At the center of the amplifier band, the gain of a device with an effective signal resistance $R_{sq}<0$ is $G^{1/2}=\left(-|R_{sq}|-Z_0\right)/\left(-|R_{sq}|+Z_0\right)=1/\Gamma_p$, where $\Gamma_p$ is the reflection coefficient of an identical circuit having a passive, positive resistance load $|R_{sq}|$.\cite{MYJ} Therefore, the problem of designing the gain of a parametric amplifier can be mapped onto the problem of impedance-matching a passive load of equal magnitude.  This also implies that the gain-bandwidth product of a parametric amplifier is only limited by the Bode-Fano theorem.\cite{Fano50} If the pumped nonlinearity in a JPA has an effective admittance $Y_{sq}(\omega_0)$, then the bandwidth $\Delta\omega$ and the power gain $G$ are related via $\Delta\omega\times\ln G^{1/2}\leq -\pi\omega_0|Re\left\{Y_{sq}(\omega_0)\right\}|/Im\left\{Y_{sq}(\omega_0)\right\}\sim\pi\omega_0^2L_a/|R_{sq}|$, where $L_a$ is the linear inductance of the junction (or SQUID) that depends on the flux bias, $\Phi_{dc}$, and $\omega_0$ is the amplifier center frequency. For typical parameters, a 7.5 GHz amplifier with 25 dB of gain, a linear inductance $L_a=90$ pH, and an effective negative resistance of $R_{sq}=1/Re\left\{Y_{sq}(\omega_0)\right\}=$ -10 $\Omega$, could have a maximum bandwidth of about 3.5 GHz. With a three-pole physically realizable matching network, one can theoretically achieve up to 60\% of that bandwidth.\cite{Kuh61}  

\begin{figure}
\includegraphics[width=3.3in]{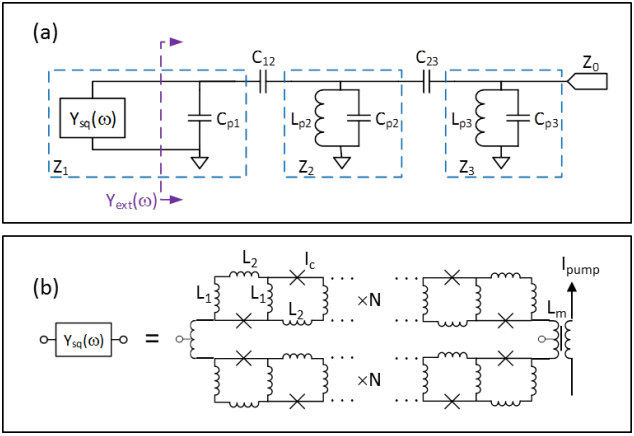}
\caption{\label{fig1} (a) Schematic of the amplifier circuit including a band-pass matching network embedding the pumped Josephson nonlinearity, represented by $Y_{sq}$. The signal port is shown on the right and labeled $Z_0$. $Y_{ext}$ is the admittance seen by the nonlinearity out through the matching network. (b) Schematic of the particular implementation of the active element in our amplifier, built with two rf-SQUID arrays in parallel. Each array has $N$ sections that contain a junction with critical current $I_c$ shunted by inductors $L_1$ and $L_2$. The pump is coupled inductively to the loop and the self inductance of the coupling transformer, $L_m$ is treated as a parasitic in our calculation.}
\end{figure}

\begin{figure}
\includegraphics[width=3.3in]{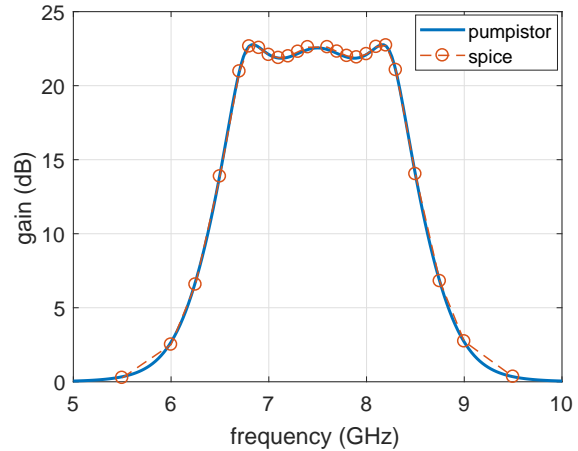}
\caption{\label{fig2} Amplifier gain as a function of signal frequency, as calculated using the pumpistor model and the circuit parameters given in the text (solid), and as simulated in WRSpice (circles).}
\end{figure}

In what follows, we will use the so-called pumpistor model \cite{Sundqvist14} to obtain an expression for $Y_{sq}(\omega)$, and design a three-pole bandpass network to match the effective load to 50 $\Omega$. Figure \ref{fig1}(a) shows the overall topology of the resulting circuit, whose gain characetristics are shown in Fig.\ \ref{fig2}. We will outline a design procedure that is quite general to Josephson amplifiers, however, we will concentrate on a particular nonlinear element shown in Figure \ref{fig1}(b) \textemdash a parallel arrangement of two rf-SQUID arrays,\cite{Zhang17} which we have previously demonstrated to be capable of carrying up to -53 dBm of power,\cite{Naaman17} and should therefore allow amplification of up to -90 dBm input signals with 20 dB of gain without saturation. The relatively high junction $I_c$ in the array can support higher mode currents than a typical JPA, and the low-inductance shunt of each of the junctions eliminates phase slips that are likely to occur otherwise when the amplifier is biased and pumped.  

The design of the matching network follows Ref.\ \onlinecite{MYJ}. We choose to implement a three-pole coupled-resonator bandpass network; larger number of poles gives only marginal improvement in bandwidth per pole. The set of filter prototype coefficients $\left\{g_i\right\}$ that describe the network, which were calculated specifically to accommodate negative-resistance loads, are tabulated in Refs.\ \onlinecite{MYJ} and \onlinecite{Getsinger63} for specified gain and ripple characteristics. Once the center frequency of the network (which we take to coincide with half the pump frequency $\omega_0=\omega_p/2$) is specified, as well as its desired bandwidth, gain, and ripple, the only remaining free parameter in the design is the capacitance shunting the Josephson element. All other parameters, particularly the dc flux bias to the Josephson element and the amplitude of the pump tone, will be constrained and can be calculated from the design equations.

The first pole of the network, resonator $Z_1$ in Fig.\ \ref{fig1}(a), is comprised of the Josephson element's linear inductance $L_a(\Phi_{dc})$ in parallel with a shunt capacitance $C_1$. The choices for $C_1$ and $\omega_0$ determine both the required flux bias $\Phi_{dc}$ operating point such that $L_a(\Phi_{dc})=1/\omega_0^2C_1$, as well as the impedance of the resonator, $Z_1=\sqrt{L_a(\Phi_{dc})/C_1}$. The other two poles of the network are built with passive $LC$ resonators whose frequencies are $\omega_0$, and impedances are $Z_2=\sqrt{L_{p2}/C_2}$ and $Z_3=\sqrt{L_{p3}/C_3}$ (the values of the resonator shunt capacitances will be modified below from $C_{1\ldots3}$ to $C_{p1\ldots3}$). The resonators are coupled via admittance inverters $J_{ij}$ whose values are calculated from $\left\{g_i\right\}$ according to $J_{12}=w/\sqrt{Z_1Z_2g_1g_2}$, $J_{23}=w/\sqrt{Z_2Z_3g_2g_3}$, and $J_{34}=\sqrt{w/\left(Z_0Z_3g_3g_4\right)}$, where $w$ is the fractional bandwith, $Z_0=50~\Omega$, and with the additional constraint\cite{MYJ}
\begin{equation}\label{bw}
  w\times\frac{|R_{sq}|}{Z_1}=g_1.
\end{equation}

We choose the impedance of the passive resonator $Z_3$ to satisfy $Z_3=wZ_0/g_3g_4$; this allows us to eliminate the last inverter $J_{34}$. The impedance $Z_2$ can be chosen arbitrarily, and we set $Z_2=\sqrt{Z_1Z_3}$. We can now implement all admittance inverters as capacitive pi-sections \cite{PozarBook} with $C_{ij}=J_{ij}/\omega_0$ to obtain the circuit shown in Fig.\ \ref{fig1}(a), where $C_{p1}=C_1-C_{12}$, $C_{p2}=C_2-C_{12}-C_{23}$ and $C_{p3}=C_3-C_{23}$, and all component values are determined by the above. The admittance seen by the nonlinear element looking through the matching network out to the 50 $\Omega$ environment, $Y_{ext}(\omega)$ in Fig.\ \ref{fig1}(a), can be evaluated at the center of the band:
\begin{equation}\label{yext}
  Y_{ext}(\omega_0)=j\omega_0C_1+\left(\frac{J_{12}}{J_{23}}\right)^2\frac{1}{Z_0}.
\end{equation}

The matching network design is now complete, and without assuming anything about the particular form of the Josephson nonlinearity, it is quite general and can be used to broadband amplifiers based on dc-SQUIDs, Josephson dipole elements,\cite{Frattini17} and rf-SQUID arrays (Fig.\ \ref{fig1}(b)) alike. However, we still need to find the optimal pump amplitude $\Phi_{ac}$ and calculate the gain profile, which require knowledge of the admittance $Y_{sq}$. We find this admittance by use of the pumpistor model of Ref.\ \onlinecite{Sundqvist14} for a flux-pumped nonlinearity in 3-wave mixing operation, which we can write as $Y_{sq}(\omega_s)=1/j\omega_sL_a+1/j\omega_s(L_b+L_c)$, where $\omega_s$ is the signal frequency, $L_a=L_T(\Phi_{dc})$ is the linear inductance of the Josephson element at the operating point, and
\begin{equation}\label{Lb}
  L_b=-\frac{4L_T^3(\Phi_{dc})}{L_T^\prime(\Phi_{dc})^2}\frac{1}{\Phi_{ac}^2}
\end{equation}
\begin{equation}\label{Lc}
  L_c=\frac{4i\omega_iL_T^4(\Phi_{dc})Y_{ext}^\ast(\omega_i)}{L_T^\prime(\Phi_{dc})^2}\frac{1}{\Phi_{ac}^2},
\end{equation}
where $L_T^\prime(\Phi_{dc})$ is the flux derivative of the inductance evaluated at the operating point, and $\omega_i=\omega_p-\omega_s$ is the idler frequency. If the amplifier is built using a simple dc-SQUID with a total critical current $I_c$, then $L_T(\Phi)=\hbar/2eI_c\cos\left(\pi\Phi/\Phi_0\right)$. Our nonlinear element, shown in Fig.\ \ref{fig1}(b), is constructed from two arrays of rf-SQUIDs, each of the array's $N$ stages composed of a junction with critical current $I_c$ shunted by linear inductors $L_1$ and $L_2$. In this case we have for the two arrays in parallel\cite{Naaman17}
\begin{equation}\label{Lt} L_T\left(\delta_0(\Phi_{dc})\right)=\frac{N}{2}\frac{\left(L_1+L_2\right)L_J+L_1L_2\cos\delta_0}{L_J+\left(4L_1+L_2\right)\cos\delta_0},
\end{equation}
\begin{widetext}
\begin{equation}\label{ltprime} L_T^\prime\left(\delta_0(\Phi_{dc})\right)=\frac{\left(2L_1+L_2\right)^3L_J^2\pi\sin\delta_0}{2\Phi_0\left[\left(L_1+L_2\right)L_J+L_1L_2\cos\delta_0\right]\left[L_J+\left(4L_1+L_2\right)\cos\delta_0\right]^2},
\end{equation}
\end{widetext}
where $L_J=\hbar/2eI_c$, $\Phi_0$ is the flux quantum, and $\delta_0(\Phi_{dc})$ is given implicitely by
\begin{equation}\label{delta0} \left(\frac{1}{L_1}+\frac{1}{L_2}\right)\delta_0+\frac{1}{L_J}\sin\delta_0=\frac{\pi\Phi_{dc}}{N\Phi_0}\left(\frac{1}{L_1}+\frac{2}{L_2}\right).
\end{equation}
Using Eqs.\ (\ref{yext})-(\ref{delta0}), we can find the pump amplitude $\Phi_{ac}$ for which $R_{sq}=1/Re\left\{Y_{sq}(\omega_0)\right\}$ satisfies the constraint in Eq.\ (\ref{bw}) at the center of the band.

We now have all circuit element values, the dc flux operating point and the optimal amplitude of the pump. To calculate the gain profile of the amplifier, we have to evaluate, at each signal frequency $\omega_s$, the admittance at the idler frequency $Y_{ext}(\omega_p-\omega_s)$, and from it calculate the admittance of the pumped nonlinearity $Y_{sq}(\omega_s)$. We then calculate the impedance of the whole amplifier as seen from the 50 $\Omega$ signal port to the right of Fig.\ \ref{fig1}(a), $Z_{amp}(\omega_s)$, and the gain in dB is given by
\begin{equation}\label{gain}
  G(\omega_s)=20\times\log_{10}\left|\frac{Z_{amp}(\omega_s)-Z_0}{Z_{amp}(\omega_s)+Z_0}\right|.
\end{equation}

As a concrete example, we design an amplifier with a center frequency of 7.5 GHz. We target a design with a gain of 25 dB and with 0.5 dB gain ripple. From tables in Refs.\ \onlinecite{MYJ},\onlinecite{Getsinger63}, we find the coefficients $g_1=0.6068$, $g_2=0.6742$, $g_3=0.3836$, and $g_4=0.8992$. We choose $w=0.25$ for the fractional bandwidth parameter, and an initial shunt capacitance of $C_1=5.2$ pF. Each rf-SQUID array is implemented with 20 Josephson junctions with $I_c=35~\mu$A, and the array inductors are $L_1=1.45$ pH and $L_2=3.52$ pH. The loop enclosing the two arrays is coupled to the pump via a 10 pH transformer with a self inductance of $L_m=20$ pH. Following the above procedure we find the dc flux operating point $\Phi_{dc}=0.26~\Phi_0$ per junction (total of 10.38 $\Phi_0$ over the entire loop of $2N=40$ junctions), and the pump amplitude that gives $R_{sq}=-9.15~\Omega$ at $\omega_0$ and satisfies Eq.\ (\ref{bw}) is found to be $\Phi_{ac}=0.073~\Phi_0$ per junction for a total of 2.93 $\Phi_0$ over the entire 40-junction loop. With reference to Fig.\ \ref{fig1}(a), the coupling capacitors evaluate to $C_{12}=1.19$ pF and $C_{23}=0.5$ pF, and the final shunt capacitors are $C_{p1}=4.01$ pF, $C_{p2}=C_{p3}=85$ fF. The passive resonator inductances are $L_{p2}=254$ pH, and $L_{p3}=769$ pH.

The solid curve in Figure \ref{fig2} shows the resulting gain characteristics of the amplifier as calculated using Eq.\ (\ref{gain}) and the pumpistor model with Eq.\ (\ref{Lb})-(\ref{delta0}). We see that the amplifier has a maximum gain of 22.8 dB, a 3 dB bandwidth of 1.63 GHz, and gain ripple of 0.74 dB. We have additionally simulated the full nonlinear circuit in WRSpice \cite{Whiteley91} in the time domain using transient analysis. We extracted the reflection coefficient of the circuit from the voltage waveform at the port of the amplifer by numerical I-Q demodulation, allowing us to separate the reflected waves from the incident signal. The results of these simulations are shown as circles in Fig.\ \ref{fig2}, and are in excellent agreement with the calculated response. In the simulation, we found that the optimal pump amplitude and dc flux were both higher than estimated within the pumpistor model by roughly 20\% and 30\% respectively, which is not surprising as the pumpistor is a linearized model while the Spice simulation captures the full nonlinearity of the circuit.

\begin{figure}
\includegraphics[width=3.3in]{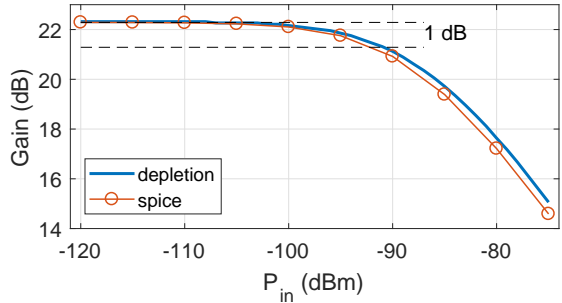}
\caption{\label{fig3} Loaded gain as a function of input signal power at $\omega_s/2\pi=7.3$ GHz from Spice simulation (circles), and from a calculation using the pumpistor model with a pump depletion saturation mechanism (solid). The simulation and calculation yield 1 dB input compression powers of -92.2 dBm and -91 dBm, respectively.}
\end{figure}

In Figure \ref{fig3}, we use the Spice simulation to characterize the large-signal response of the circuit and estimate its 1 dB compression power, independent of any theory for the saturation mechanism. The simulated data are shown in the figure as circles for a signal frequency of 7.3 GHz, and we obtain a saturation power of $P_{sat}=-92.2$ dBm. We compare the simulated data to a calculation within the pumpistor model framework, in which we assume a pump depletion saturation mechanism, and which is shown in Fig.\ \ref{fig3} as solid line. Our calculation is based on a power balance equation for the pump, recognizing that the conversion of pump photon into signal and idler photons present the pump mode with an effective loss channel.\cite{Abdo13} The loaded pump power $P_{load}$ available to drive the amplification process is $P_{load}=P_{av}-P_{loss}$, where  $P_{av}$ is the available power from the pump port, and $P_{loss}=\frac{1}{2} G_{load}\left(2P_{in}+w\hbar\omega_0^2\right)$. Here the last term is the input quantum noise power over the amplifier band, $P_{in}$ is the input signal power, and $G_{load}$ is the loaded gain evaluated with $P_{load}$ drive. The available and loaded pump powers relate to the respective pump amplitudes $\Phi_{ac}$ such that $P_{av}=\omega_p\Phi_{ac,av}^2/8L_T$ and similarly for the loaded amplitude. The solid curve in Fig.\ \ref{fig3} is a plot of $G_{load}$ and is calculated by solving self consistently for the loaded gain,
\begin{equation}\label{sat}
  \Phi_{ac,load}^2=\Phi_{ac,av}^2-\frac{4L_T}{\omega_p}G_{load}\left(2P_{in}+w\hbar\omega_0^2\right).
\end{equation}

The pump depletion model gives $P_{sat}=-91$ dBm, and is in reasonable agreement with the simulated data. We recognize that other saturation mechanism exist,\cite{Sundqvist14,Abdo13,Liu17} and while Fig.\ \ref{fig3} is not sufficient evidence that pump depletion dominates, the figure does suggest that competing mechanisms contribute up to the same order in this type of amplifier.

To summarize, we have shown how to design an impedance matching network to embed Josephson parametric amplifiers and achieve broadband performance to meet prescribed gain, bandwidth, and ripple specifications. We additionally introduced a Josephson element composed of rf-SQUID arrays to achieve high saturation powers. Calculated gain and saturation curves agree well with simulated data that capture the full nonlinear circuit. Similar techniques can be readily implemented to broadband amplifiers based on the JPC architecture, but will require separate designs for the spatially and spectrally distinct signal and idler modes.

We thank D. Dawson and A. Marakov for technical assistance.

%\section{}
%\label{}
%\subsection{}
%\subsubsection{}

% If in two-column mode, this environment will change to single-column format so that long equations can be displayed. 
%\begin{widetext}
%$$\mbox{put long equation here}$$
%\end{widetext}

% Figures should be put into the text as floats. 
% Use the graphics or graphicx packages (distributed with LaTeX2e).
% Put the label that you will use with \ref{} command in the braces of the \label{} command.
%
% \begin{figure}
% \end{figure}

% \begin{table}
% \end{tabular}
% \end{table}

%\begin{acknowledgments}
% Put your acknowledgments here.
%\end{acknowledgments}

\bibliography{paramp_APL}

\end{document}